\documentclass[11pt]{article}

\usepackage{amssymb}
\usepackage{amsmath}
\usepackage{amsfonts}

\newcommand{\R}{\mathbb{R}}
\newcommand{\C}{\mathbb{C}}
\newcommand{\Z}{\mathbb{Z}}

\newcommand{\be}{\begin{equation}}
\newcommand{\bea}{\begin{eqnarray}}
\newcommand{\eea}{\end{eqnarray}}
\newcommand{\nn}{\nonumber}
\newcommand{\kt}{\rangle}
\newcommand{\br}{\langle}

\newcommand{\qq}{{\cal Q}}
\newcommand{\ed}{\end{document}}
\newcommand{\bbr}{\br\!\br}
\newcommand{\kkt}{\kt\!\kt}

\begin{document}

\title{Pseudo-Hermiticity versus $PT$ Symmetry: The necessary condition for the reality of the
spectrum of a non-Hermitian Hamiltonian}
\author{Ali Mostafazadeh\thanks{E-mail address:
amostafazadeh@ku.edu.tr}\\ \\
Department of Mathematics, Ko\c{c} University,\\
Rumelifeneri Yolu, 80910 Sariyer, Istanbul, Turkey}
\date{ }
\maketitle

\begin{abstract}
We introduce the notion of {\em pseudo-Hermiticity} and show that
every Hamiltonian with a real spectrum is pseudo-Hermitian. We
point out that all the $PT$-symmetric non-Hermitian Hamiltonians
studied in the literature belong to the class of pseudo-Hermitian
Hamiltonians, and argue that the basic structure responsible for
the particular spectral properties of these Hamiltonians is their
pseudo-Hermiticity. We explore the basic properties of general
pseudo-Hermitian Hamiltonians, develop {\em pseudo-supersymmetric
quantum mechanics}, and study some concrete examples, namely the
Hamiltonian of the two-component Wheeler-DeWitt equation for the
FRW-models coupled to a real massive scalar field and a class of
pseudo-Hermitian Hamiltonians with a real spectrum.
\end{abstract}

\baselineskip=24pt

\section{Introduction}

The past three years have witnessed a growing interest in
non-Hermitian Hamiltonians with real spectra \cite{bb-98} -
\cite{kr-sz}. Based on the results of various numerical studies,
Bender and his collaborators \cite{bb-98,bbm-99} found certain
examples of one-dimensional non-Hermitian Hamiltonians that
possessed real spectra. Because these Hamiltonians were invariant
under $PT$ transformations, their spectral properties were linked
with their $PT$-symmetry. The purpose of this article is to
explore the basic structure responsible for the reality of the
spectrum of a non-Hermitian Hamiltonian.

By definition, a $PT$-symmetric Hamiltonian $H$ satisfies
    \be
    PT H (PT)^{-1}=PT H PT  = H,
    \label{pt}
    \end{equation}
where $P$ and $T$ are respectively the operators of parity and time-reversal transformations.
These are defined according to
    \be
    P\, x\, P=-x,~~~~P\,p\,P=T\,p\,T=-p,~~~~ T\, i1\, T=-i1,
    \label{pt-def}
    \end{equation}
where $x, p, 1$ are respectively the position, momentum, and
identity operators acting on the Hilbert space ${\cal H}=L^2(\R)$
and $i:=\sqrt{-1}$. Note that Eqs.~(\ref{pt-def}) apply only for
the systems whose classical position $x$ and momentum $p$ are
real. In this article we shall only be concerned with these
systems.

As we mentioned above, the only reason for relating the concept of
$PT$-symmetry and non-Hermitian Hamiltonians with a real spectrum
is that most of the known examples of the latter satisfy
Eq.~(\ref{pt}). Certainly there are Hermitian Hamiltonians with a
real spectrum that are not $PT$-symmetric and there are
$PT$-symmetric Hamiltonians that do not have a real spectrum.
Therefore, $PT$-symmetry is neither a necessary nor a sufficient
condition for a Hamiltonian to have a real spectrum. This raises
the possibility that the $PT$-symmetry of a Hamiltonian may have
nothing to do with the reality of its spectrum. The interest in
$PT$-symmetry seems to be mostly because of the lack of an
alternative framework replacing the Hermiticity of the Hamiltonian
in ordinary (unitary) quantum mechanics. Much of the published
work on the subject concerns the study of various examples and the
extension of the concepts developed for Hermitian Hamiltonians to
the $PT$-symmetric ones, \cite{bb-98} - \cite{z-01}. Recently,
Znojil \cite{z-april-01}, Japaridze \cite{japaridze}, Kretschmer
and Szymanowski \cite{kr-sz} have addressed some of the more
fundamental issues regarding the mathematical structure and the
interpretation of the $PT$-symmetric quantum mechanics.

Among the common properties of all the $PT$-symmetric Hamiltonians that have so far been studied are the
following.
    \begin{itemize}
    \item[1.] Either the spectrum of the Hamiltonian is real ($PT$-symmetry is exact) or there are
    complex-conjugate pairs of complex eigenvalues ($PT$-symmetry is broken), \cite{bb-98,bbm-99,dt-00,km-00};
    \item[2.] The indefinite inner-product $\bbr~|~\kkt$ defined by
        \be
        \bbr\psi_1|\psi_2\kkt:=\br\psi_1|P|\psi_2\kt,~~~~~\forall |\psi_1\kt,|\psi_2\kt\in{\cal H},
        \label{P-braket}
        \end{equation}
    is invariant under the time-translation generated by the Hamiltonian, \cite{z-april-01,japaridze}.
    \end{itemize}
The main motivation for the present investigation is the
remarkable fact that there is no evidence that $PT$-symmetry is
the basic structure responsible for these properties. For example,
in Ref.~\cite{cjt-98}, the authors construct a class of
non-$PT$-symmetric Hamiltonians with a real spectrum. Another
example of a non-Hermitian Hamiltonian with similar properties is
the Hamiltonian describing the evolution of the solutions of the
two-component Wheeler-DeWitt equation for FRW-models coupled with
a real massive scalar field \cite{jmp-98}. This Hamiltonian is
explicitly `time-dependent,' `parity-invariant,' and non-Hermitian
(with respect to the relevant $L^2-norm$ on the space of
two-component wave functions), but the corresponding invariant
indefinite inner-product does not involve $P$.

The organization of the article is as follows. In Section~2, we introduce the concept of a
{\em pseudo-Hermitian} operator and derive the basic spectral properties of pseudo-Hermitian
Hamiltonians. These coincide with Properties~1 and~2 (with $P$ replaced with a Hermitian invertible
linear operator $\eta$). In section~3, we consider the class of pseudo-Hermitian Hamiltonians that
have a complete biorthonormal eigenbasis and show that the pseudo-Hermiticity is a necessary
condition for having a real spectrum. In Section~4, we explore the pseudo-Hermitian Hamiltonian of
the two-component Wheeler-DeWitt equation for FRW-models coupled with a real massive scalar field.
In Section~5, we develop pseudo-supersymmetric quantum mechanics. In Section~6, we use
pseudo-supersymmetry to construct a large class of pseudo-Hermitian Hamiltonians with a real
spectrum.  In Section~7, we present our concluding remarks.

\section{Pseudo-Hermitian Hamiltonians}

We first give a few definitions. Throughout this paper we will assume that all the inner product spaces
are complex. The generalization to real inner product spaces is straightforward.
\begin{itemize}
\item[~] {\bf Definition~1:}
Let $V_\pm$ be two inner product spaces endowed with Hermitian linear automorphisms $\eta_\pm$
(invertible operators mapping $V_\pm$ to itself and satisfying
    \[ \forall v_\pm,w_\pm\in V_\pm,~~~(v_\pm,\eta_\pm w_\pm)_\pm=(\eta_\pm v_\pm,w_\pm)_\pm,\]
where $(~,~)_\pm$ stands for the inner product of $V_\pm$) and $O:V_+\to V_-$ be a linear operator. Then
the $\eta_\pm$-pseudo-Hermitian adjoint $O^\sharp: V_-\to V_+$ of $O$ is defined by
$O^\sharp:=\eta_+^{-1}O^\dagger\eta_-$. In particular, for $V_\pm=V$ and $\eta_\pm=\eta$, the operator $O$
is said to be $\eta$-pseudo-Hermitian if $O^\sharp=O$.
\item[~] {\bf Definition~2:}
Let $V$ be an inner product space. Then a linear operator $O:V\to V$ is said to be pseudo-Hermitian, if
there is a Hermitian linear automorphism $\eta$ such that $O$ is $\eta$-pseudo-Hermitian.
\end{itemize}

Now, consider a quantum system with a possibly non-Hermitian and time-dependent Hamiltonian $H=H(t)$ and
a Hilbert space ${\cal H}$ which is endowed with a Hermitian linear automorphism $\eta$.
\begin{itemize}
\item[~]{\bf Proposition~1:}  The Hermitian indefinite inner product $\bbr~|~\kkt_\eta$ defined by $\eta$, i.e.,
    \be
    \bbr\psi_1|\psi_2\kkt_\eta :=\br\psi_1|\eta|\psi_2\kt,~~~~~\forall |\psi_1\kt,|\psi_2\kt\in{\cal H},
    \label{indef}
    \end{equation}
is invariant under the time-translation generated by the Hamiltonian $H$ if and only if $H$ is
$\eta$-pseudo-Hermitian.
\item[~]{\bf Proof:} First note that the $\eta$-pseudo-Hermiticity of $H$ is equivalent to the condition
    \be
    H^\dagger=\eta\, H\,\eta^{-1}.
    \label{p-hermitian}
    \end{equation}
Now, using the Schr\"odinger equation
    \be
    i\frac{d}{dt}\, |\psi(t)\kt=H|\psi(t)\kt\;,
    \label{sch-eq}
    \end{equation}
its adjoint, and Eq.~(\ref{indef}), one has for any two evolving state vectors $|\psi_1(t)\kt$ and $|\psi_2(t)\kt$:
    \[ i\frac{d}{dt} \bbr\psi_1(t)|\psi_2(t)\kkt_\eta=\br\psi_1|(\eta H- H^\dagger\eta) |\psi_2\kt.\]
Therefore, $\bbr\psi_1(t)|\psi_2(t)\kkt_\eta$ is a constant if and only if (\ref{p-hermitian}) holds.~$\square$
\end{itemize}

Note that choosing $\eta=1$ reduces Eq.~(\ref{p-hermitian}) to the
condition of the Hermiticity of the Hamiltonian. Hence {\em
pseudo-Hermiticity is a generalization of Hermiticity.}
Furthermore, observe that a typical $PT$-symmetric Hamiltonian
defined on a real phase space ($(x,p)\in\R^2$) has the form
$H=p^2/(2m)+V(x)$ where the potential $V(x)=V_+(x) +i V_-(x)$ has
an even real part $V_+(x)$ and an odd imaginary part $V_-(x)$,
i.e., $V_\pm(\pm x)=\pm V_\pm(x)$. It is not difficult to see that
such a $PT$-symmetric Hamiltonian satisfies
    \[H^\dagger= \frac{p^2}{2m}+V_+(x)-iV_-(x) = \frac{p^2}{2m}+V_+(-x)+iV_-(-x) = P\,H\,P=P\,H\,P^{-1}.\]
Hence it is $P$-pseudo-Hermitian. In contrast, consider the non-Hermitian Hamiltonians
    \[H_1:=p^2+x^2p,~~~~~H_2:=p^2+i(x^2p+p\,x^2).\]
Clearly, $H_1$ is $PT$ symmetric, but not $P$-pseudo-Hermitian.
Whereas, $H_2$ is $P$-pseudo-Hermitian and not $PT$ symmetric.
Therefore, $PT$ symmetry and $P$-pseudo-Hermiticity are distinct
properties. Note, however, that $H_1$ may be pseudo-Hermitian with
respect to another Hermitian automorphism $\eta$. We shall explore
the relationship between $PT$-symmetry and pseudo-Hermiticity in
the next section.

The defining condition~(\ref{p-hermitian}) may also be expressed as the intertwining relation
    \be
    \eta\, H=H^\dagger\, \eta.
    \label{p-her-2}
    \end{equation}
Using this equation together with the eigenvalue equation for the Hamiltonian, namely $H|E_i\kt=E_i|E_i\kt$,
and its adjoint, we can easily show that any two eigenvectors $|E_i\kt$ and $|E_j\kt$ of $H$ satisfy
    \be
    (E_i^*-E_j) \bbr E_i|E_j\kkt_\eta=0.
    \label{eta-orth-1}
    \end{equation}
A direct implication of this equation is the following Proposition.
    \begin{itemize}
    \item[~] {\bf Proposition~2:} An $\eta$-pseudo-Hermitian Hamiltonian has the following properties.
        \begin{itemize}
        \item[~](a) The eigenvectors with a non-real eigenvalue have vanishing $\eta$-semi-norm, i.e.,
            \be
            E_i\notin\R~~~~{\rm implies}~~~~|\!|\, |E_i\kt |\!|_\eta^2:=\bbr E_i|E_i\kkt_\eta=0;
            \label{s1}
            \end{equation}
        \item[~](b) Any two eigenvectors are $\eta$-orthogonal unless their eigenvalues are complex         conjugates, i.e.,
            \be
            E_i\ne E_j^*~~~~{\rm implies}~~~~\bbr E_i|E_j\kkt_\eta=0.
            \label{s2}
            \end{equation}
    In particular, the eigenvectors with distinct real eigenvalues are $\eta$-orthogonal.
    \end{itemize}
    \end{itemize}

In the remainder of this section, we list a number of simple but remarkable consequences of
pseudo-Hermiticity.
    \begin{itemize}
    \item[~] {\bf Proposition~3:}  Let $V$ be an inner product space endowed with a Hermitian linear
        automorphism $\eta$, $1:V\to V$ denote the identity operator, $O_1,O_2:V\to V$ be linear
        operators, and $z_1,z_2\in\C$. Then,
            \begin{itemize}
            \item[~] (a) $1^\sharp=1$;
            \item[~] (b) $(O_1^\sharp)^\sharp=O_1$;
            \item[~] (c) $(z_1 O_1+z_2 O_2)^\sharp=z_1^* O_1^\sharp+z_2^* O_2^\sharp$,
            \end{itemize}
        where $z_i^*$ stands for the complex conjugate of $z_i$.
    \item[~] {\bf Proof:} (a) and (b) are trivial consequences of the definition of $\sharp$ and the
        Hermiticity of $\eta$. (c) follows from this definition and the linearity of $\eta$ and $\eta^{-1}$:
            \[(z_1 O_1+z_2 O_2)^\sharp=\eta^{-1}(z_1 O_1+z_2 O_2)^\dagger\eta=
            z_1^*\eta^{-1}O_1^\dagger\eta+z_2^*\eta^{-1}O_2^\dagger\eta=z_1^* O_1^\sharp+
            z_2^* O_2^\sharp.~\square\]
    \item[~] {\bf Proposition~4:} Let $V_\ell$, with $\ell\in\{1,2,3\}$, be inner product spaces
        endowed with Hermitian linear automorphisms $\eta_\ell$ and $O_1:V_1\to V_2$ and
        $O_2:V_2\to V_3$ be linear operators. Then $(O_2 O_1)^\sharp =O_1^\sharp O_2^\sharp$.
    \item[~] {\bf Proof:} This relation follows from the following simple calculation.
            \[(O_2 O_1)^\sharp=\eta_1^{-1}(O_2O_1)^\dagger\eta_3=
            \eta_1^{-1}O_1^\dagger\eta_2\eta_2^{-1} O_2^\dagger\eta_3=O_1^\sharp O_2^\sharp.
            ~\square\]
    \item[~] {\bf Corollary:} Pseudo-Hermitian conjugation ($O\to O^\sharp$) is a $*$-operation.
    \item[~] {\bf Proof:} According to Prop.~3 and Prop.~4, $\sharp$ has all the properties of a
        $*$-operation.~$\square$
    \item[~] {\bf Proposition~5:} Let $V$ be an inner product space endowed with a Hermitian linear
    automorphism $\eta$, $U:V\to V$ be a unitary operator, and $O:V\to V$ be a linear operator.
    Then $\eta_U:=U^\dagger\eta U$ is a Hermitian linear automorphism, and $O$ is
    $\eta$-pseudo-Hermitian if and only if $O_U:=U^\dagger OU$ is $\eta_U$-pseudo-Hermitian.
    In other words, the notion of pseudo-Hermiticity is unitary-invariant.
    \item[~] {\bf Proof:} First we recall that because $U$ is unitary, $\eta_U$ is both
    Hermitian and invertible. Furthermore, we have
        \[\eta_U^{-1}O_U^\dagger\eta_U=U^\dagger\eta^{-1} U
        U^\dagger O^\dagger U U^\dagger\eta U=U^\dagger(\eta^{-1}O^\dagger\eta)U.~
        \square\]
    \item[~] {\bf Proposition~6:}  Let $V$ be an inner product space, $\eta_1$ and $\eta_2$ be
    Hermitian linear automorphisms, and $O:V\to V$ be a linear operator. Then
    $\eta_1$-pseudo-Hermitian adjoint of $O$ coincide with its $\eta_2$-pseudo Hermitian adjoint
    if and only if $\eta_2^{-1}\eta_1$ commutes with $O$.
    \item[~] {\bf Proof:} This statements holds because $\eta_1^{-1}O^\dagger\eta_1=
    \eta_2^{-1}O^\dagger\eta_2$ implies $O^\dagger\eta_1\eta_2^{-1}=\eta_1\eta_2^{-1}O^\dagger$.
    Taking the Hermitian adjoint of this relation yields $[O,\eta_2^{-1}\eta_1]=0$.\;$\square$
    \item[~] {\bf Corollary:} If the Hamiltonian $H$ of a quantum system is pseudo-Hermitian with
    respect to two different Hermitian linear automorphisms $\eta_1$ and $\eta_2$ of the Hilbert
    space, then $\eta_2^{-1}\eta_1$ is a symmetry of the system. Conversely, let $\eta$ be a
    Hermitian linear automorphism of the Hilbert space, $G$ be a symmetry group of the system
    whose elements $g$ are represented by invertible linear operators. Then $\eta g$ is a Hermitian
    linear automorphism and $H$ is $\eta g$-pseudo-Hermitian provided that $g^\dagger\eta g=\eta$.
    \item[~] {\bf Proof:} This is a direct implication of Prop.~6 and the definition of the
    symmetry, namely $[g,H]=0$ or equivalently $g^{-1}Hg=H$.
    \end{itemize}

\section{Pseudo-Hermitian Hamiltonians with a Complete Biorthonornal Eigenbasis}

Let $H$ be an $\eta$-pseudo-Hermitian Hamiltonian with a complete biorthonormal eigenbasis
$\{|\psi_n,a\kt,|\phi_n,a\kt\}$ and a discrete spectrum, \cite{biorthonormal}. Then, by definition,
    \bea
    &&H|\psi_n,a\kt=E_n|\psi_n,a\kt,~~~~H^\dagger|\phi_n,a\kt=E_n^*|\phi_n,a\kt\;,
    \label{o1}\\
    &&\br\phi_m,b|\psi_n,a\kt=\delta_{mn}\delta_{ab},
    \label{o2}\\
    && \sum_n\sum_{a=1}^{d_n}|\phi_n,a\kt\br\psi_n,a| =\sum_n\sum_{a=1}^{d_n}|\psi_n,a\kt\br\phi_n,a|=1,
    \label{o3}
    \eea
where $d_n$ is the multiplicity (degree of degeneracy) of the eigenvalue $E_n$, and $a$ and $b$ are
degeneracy labels.
\begin{itemize}
\item[~] {\bf Proposition~7:} Let $H$ be a pseudo-Hermitian Hamiltonian with these properties.
Then the non-real eigenvalues of $H$ come in complex conjugate
pairs with the same multiplicity.
\item[~] {\bf Proof:} According to Eqs.~(\ref{p-hermitian}) and (\ref{o1}),
    \be
    H\left(\eta^{-1}|\phi_n,a\kt\right)=\eta^{-1}H^\dagger|\phi_n,a\kt=
    E_n^*\left(\eta^{-1}|\phi_n,a\kt\right).
    \label{o4}
    \end{equation}
Because $\eta^{-1}$ is invertible, $\eta^{-1}|\phi_n,a\kt\ne 0$ is an eigenvector of $H$ with eigenvalue
$E_n^*$. More generally, $\eta^{-1}$ maps the eigensubspace associated with $E_n$ to the that associated
with $E_n^*$. Again, because $\eta^{-1}$ is invertible, $E_n$ and $E_n^*$ have the same
multiplicity.~$\square$
\end{itemize}

Next, we use the subscript `$~_0$' to denote real eigenvalues and the corresponding basis eigenvectors and
the subscript `$~_\pm$' to denote the complex eigenvalues with $\pm$ imaginary part and the corresponding
basis eigenvectors. Then  in view of Eqs.~(\ref{o1}) -- (\ref{o4}), we have
    \bea
    1&=& \sum_{n_0}\sum_{a=1}^{d_{n_0}} | \psi_{n_0},a \kt \br \phi_{n_0},a| +
    \sum_{n_+}\sum_{\alpha=1}^{d_{n_+}} \left( | \psi_{n_+},\alpha \kt \br \phi_{n_+},\alpha| +
    | \psi_{n_-},\alpha \kt \br \phi_{n_-},\alpha|\right),
    \label{1=}\\
    H&=& \sum_{n_0}\sum_{a=1}^{d_{n_0}} E_{n_0} | \psi_{n_0},a \kt \br \phi_{n_0},a| +
    \sum_{n_+}\sum_{\alpha=1}^{d_{n_+}} \left( E_{n_+}  | \psi_{n_+},\alpha \kt \br \phi_{n_+},\alpha| +
    E_{n_+}^*  | \psi_{n_-},\alpha \kt \br \phi_{n_-},\alpha|\right),
    \label{H=}
    \eea

Repeating the calculation leading to Eq.~(\ref{o4}), we find
    \bea
    \eta^{-1}|\phi_{n_0},a\kt &=&\sum_{b=1}^{d_{n_0}}c^{(n_0)}_{ba} |\psi_{n_0},b\kt ,~~~~
    c^{(n_0)}_{ab}:=\br\phi_{n_0},a|\eta^{-1}|\phi_{n_0},b\kt,
    \label{e1}\\
    \eta^{-1} |\phi_{n_+},\alpha\kt &=&\sum_{\beta=1}^{d_{n_+}}c^{(n_+)}_{\beta\alpha}|\psi_{n_-},
    \beta\kt,~~~~   c^{(n_+)}_{\alpha\beta}:=\br\phi_{n_-},\alpha|\eta^{-1}|\phi_{n_+},\beta\kt,
    \label{e2}\\
    \eta^{-1}|\phi_{n_-},\alpha\kt&=&\sum_{\beta=1}^{d_{n_+}}c^{(n_-)}_{\beta\alpha}|\psi_{n_+},
    \beta\kt,~~~~   c^{(n_-)}_{\alpha\beta}:=\br\phi_{n_+},\alpha|\eta^{-1}|\phi_{n_-},\beta\kt,
    \label{e3}
    \eea
where $c^{(n_0)}_{ab}$ and $c^{(n_\pm)}_{\alpha\beta}$ are complex coefficients. The latter may be viewed
as entries of complex matrices $c^{(n_0)}$ and $c^{(n_\pm)}$, respectively. Because $\eta$ and consequently
$\eta^{-1}$ are Hermitian operators, so are the matrices $c^{(n_0)}$ and $c^{(n_\pm)}$. In particular, we
can make a unitary transformation of the Hilbert space to map the biorthonormal system of eigenbasis
vectors of the Hamiltonian to a new system in which these matrices are diagonal. We can further rescale
the basis vectors so that $c^{(n_0)}$ and $c^{(n_\pm)}$ become identity matrices. In the following we
shall assume, without loss of generality, that such a transformation has been performed. Then,
Eqs.~(\ref{e1}) -- (\ref{e3}) take the form
    \be
    |\phi_{n_0},a\kt=\eta |\psi_{n_0},a\kt,~~~~
    |\phi_{n_\pm},\alpha\kt=\eta|\psi_{n_\mp},\alpha\kt.
    \label{p=p}
    \end{equation}
In particular, combining this result with Eq.~(\ref{o2}), we have
the following $\eta$-orthonormalization of the eigenvectors of
$H$.
    \be
    \bbr \psi_{n_0},a|\psi_{m_0},b\kkt_\eta=\delta_{n_0,m_0}\delta_{ab},~~~~
    \bbr \psi_{n_\pm},\alpha|\psi_{m_\mp},\beta\kkt_\eta=\delta_{n_\pm,m_\mp}\delta_{\alpha\beta}.
    \label{ortho}
    \end{equation}

Next, we solve Eqs.~(\ref{p=p}) for $|\psi_{n_0}\kt$ and $|\psi_{n_\pm}\kt$ and substitute the result in
Eq.~(\ref{1=}). This leads to an explicit expression for $\eta$ that can be easily inverted to yield $\eta^{-1}$.
The result is
    \bea
    \eta= \sum_{n_0}\sum_{a=1}^{d_{n_0}} | \phi_{n_0},a \kt \br \phi_{n_0},a| +
    \sum_{n_+}\sum_{\alpha=1}^{d_{n_+}} \left( | \phi_{n_-},\alpha \kt \br \phi_{n_+},\alpha| +
    | \phi_{n_+},\alpha \kt \br \phi_{n_-},\alpha|\right),
    \label{eta=}\\
    \eta^{-1}= \sum_{n_0}\sum_{a=1}^{d_{n_0}} | \psi_{n_0},a \kt \br \psi_{n_0},a| +
    \sum_{n_+}\sum_{\alpha=1}^{d_{n_+}} \left( | \psi_{n_-},\alpha \kt \br \psi_{n_+},\alpha| +
    | \psi_{n_+},\alpha \kt \br \psi_{n_-},\alpha|\right),
    \label{eta-1=}
    \eea
One can easily check that the Hamiltonian $H$ and the operators $\eta$ and $\eta^{-1}$ as given by
Eqs.~(\ref{H=}), (\ref{eta=}), and (\ref{eta-1=}) satisfy the $\eta$-pseudo-Hermiticity condition~(\ref{p-hermitian}).

The above analysis provides the following necessary and sufficient condition for pseudo-Hermiticity.
    \begin{itemize}
    \item[~]{\bf Theorem:} Let $H$ be a non-Hermitian Hamiltonian with a discrete spectrum and a complete
biorthonormal system of eigenbasis vectors $\{|\psi_n,a\kt,|\phi_n,a\kt\}$. Then $H$ is pseudo-Hermitian if and only
if one of the following conditions hold
    \begin{itemize}
    \item[1.] The spectrum of $H$ is real;
    \item[2.] The complex eigenvalues come in complex conjugate pairs and the multiplicity of complex
    conjugate eigenvalues are the same.
    \end{itemize}
    \item[~] {\bf Proof:} We have already shown in Prop.~7 that pseudo-Hermiticity of $H$ implies at least
    one of these conditions. To prove that these conditions are sufficient for the pseudo-Hermiticity of $H$,
    we use $\{|\psi_n,a\kt,|\phi_n,a\kt\}$ to express $H$ in the form~(\ref{H=}) and construct $\eta$
    according to Eq.~(\ref{eta=}). Then, by construction, $H$ and $\eta$
    satisfy~(\ref{p-hermitian}).~$\square$
    \end{itemize}
This theorem reveals the relevance of the concept of pseudo-Hermiticity to the spectral properties of the
$PT$-symmetric Hamiltonians considered in the literature. To the best of our knowledge, an analogue of this
theorem that would apply to arbitrary $PT$-symmetric Hamiltonians does not exist. A direct implication of this
theorem is the following corollary.
    \begin{itemize}
    \item[~]{\bf Corollary~1:} Every non-Hermitian Hamiltonian with a discrete real spectrum and a complete
    biorthonormal system of eigenbasis vectors is pseudo-Hermitian.
    \end{itemize}
Note that, in general, a non-Hermitian Hamiltonian may not admit a
complete biorthonormal system of eigenvectors. The preceding
Theorem and Corollary~1 may not apply for these non-Hermitian
Hamiltonians.
    \begin{itemize}
    \item[~]{\bf Corollary~2:} Every $PT$-symmetric Hamiltonian with a discrete spectrum and a complete
    biorthonormal system of eigenbasis vectors is pseudo-Hermitian.
    \item[~]{\bf Proof:} This statement follows from the above Theorem and fact that
    the eigenvalues of every $PT$-symmetric Hamiltonian with a complete
    biorthonormal system of eigenbasis vectors come in complex
    conjugate pairs. To see this, let $|E\kt$ be an eigenvector of
    $H$ with eigenvalue $E$, i.e., $H|E\kt=E|E\kt$, and $|E\kt':=PT|E\kt$. Then
        \[
        H|E\kt'=H(PT)|E\kt=(PT)H|E\kt=(PT)E|E\kt=E^*(PT)|E\kt=E^*|E\kt',\]
    where we have made use of the linearlity of $P$ and the antilinearlity of
    $T$.~$\square$
    \end{itemize}

\section{Pseudo-Hermiticity in Minisuperspace Quantum Cosmology}

The Wheeler-DeWitt equation (with a particularly simple factor ordering prescription) for a
Freedman-Robertson-Walker (FRW) model coupled to a massive real scalar field has the form
    \be
    \left[- \frac{\partial^2}{\partial\alpha^2} + \frac{\partial^2}{\partial\phi^2} +
    \kappa\, e^{4\alpha}-m^2e^{6\alpha}\phi^2\right]\psi(\alpha,\phi)=0,
    \label{wd}
    \end{equation}
where $\alpha=\ln a$, $a$ is the scale factor, $\phi$ is the scalar field, $m$ is the mass of $\phi$, and $\kappa=-1,~0$, or $1$
depending on whether the universe is open, flat, or closed, \cite{page}. In the two-component representation developed in
Ref.~\cite{jmp-98}, this equation takes the form of the Schr\"odinger equation: $i\dot\Psi=H(\alpha)\Psi$ where a dot
stands for a derivative with respect to $\alpha$ and
    \bea
    && \Psi=\frac{1}{\sqrt 2}\left(\begin{array}{c}
    \psi+i\dot\psi\\
    \psi-i\dot\psi\end{array}\right)\,,~~~~
    H=\frac{1}{2}\left(\begin{array}{cc}
    1+{\cal D} & -1+{\cal D}\\
    1-{\cal D} & -1-{\cal D}\end{array}\right),
    \label{wd1}\\
    && {\cal D}:=-\frac{\partial^2}{\partial\phi^2}+V(\phi,\alpha),~~~~
    V(\phi,\alpha):=m^2 e^{6\alpha}\phi^2-\kappa\, e^{4\alpha}.
    \label{wd2}
    \eea
As seen from these equations ${\cal D}/2$, up to an unimportant
additive scalar, is the Hamiltonian of a `time-dependent' simple
harmonic oscillator with unit `mass' and `frequency'
$\omega=m\,e^{3\alpha}$, where $\alpha$ and $\phi$ play the roles
of time $t$ and position~$x$, respectively.

It is not difficult to check that the two-component Hamiltonian
$H$ is not Hermitian with respect to the $L^2$-inner product on
the space of two-component state vectors $\Psi$. However, its
eigenvalue problem can be solved exactly \cite{jmp-98}. For an
open or flat FRW universe ($\kappa=-1,0$) the eigenvalues of $H$
are real. For a closed FRW model, there is a range of values of
$\alpha$ for which all the eigenvalues are real. Outside this
range they come in complex conjugate imaginary pairs. This
suggests that $H$ is a pseudo-Hermitian Hamiltonian. In fact, we
can easily check that $H$ is an $\eta$-pseudo Hermitian
Hamiltonian for
    \be
    \eta=\left(\begin{array}{cc}
    1&0\\
    0&-1\end{array}\right).
    \label{wd3}
    \end{equation}
The indefinite inner product corresponding to~(\ref{wd3}) is
nothing but the Klein-Gordon inner product that is invariant under
the `time-translation' generated by $H$.

\section{Pseudo-Supersymmetric Quantum Mechanics}

The application of the ideas of supersymmetric quantum mechanics \cite{sqm} in constructing non-Hermitian
$PT$-symmetric Hamiltonians has been considered in Refs.~\cite{cjt-98,aicd-99,bcq-00,cirr-01,ddt-01} and a
formulation of $PT$-symmetric supersymmetry has been outlined in Refs.~\cite{zcbr-00,z-01}. In this section,
we develop a straightforward generalization of supersymmetric quantum mechanics that applies for
pseudo-Hermitian Hamiltonians.
    \begin{itemize}
    \item[~] {\bf Definition~3:} Consider a $\Z_2$-graded quantum system \cite{npb} with the Hilbert space
    ${\cal H}_+\oplus{\cal H}_-$ and the involution or grading operator $\tau$ satisfying
        \be
        \tau=\tau^\dagger=\tau^{-1}~~~{\rm and}~~~
        \forall |\psi_\pm\kt\in{\cal H}_\pm,~~~\tau|\psi_\pm\kt=\pm|\psi_\pm\kt.
        \label{tau}
        \end{equation}
    Let $\eta$ be an even Hermitian linear automorphism (i.e., $[\eta,\tau]=0$) and suppose that the Hamiltonian
    $H$ of the system is $\eta$-pseudo-Hermitian. Then $H$ (alternatively the system) is said to have a
    pseudo-supersymmetry generated by an odd linear operator $\qq$ (i.e.,$\{\qq,\tau\}=0$) if $H$ and
    $\qq$ satisfy the pseudo-superalgebra
        \be
        \qq^2=\qq^{\sharp 2}=0,~~~~\{\qq,\qq^\sharp\}=2H.
        \label{p-sqm}
        \end{equation}
    \end{itemize}

A simple realization of pseudo-supersymmetry is obtained using the two-component representation of the
Hilbert space where the state vectors $|\psi\kt$ are identified by the column vector
$\left(\begin{array}{c}|\psi_+\kt\\ |\psi_-\kt\end{array}\right)$ of their components $|\psi_\pm\kt$ belonging to
${\cal H}_\pm$. In this representation, one can satisfy the $\eta$-pseudo-Hermiticity of the Hamiltonian $H$,
(i.e., Eq.~(\ref{p-hermitian})) and the pseudo-superalgebra (\ref{p-sqm}) by setting
    \bea
    \tau &=&\left(\begin{array}{cc}
    1& 0\\
    0 & -1 \end{array}\right),~~~~
    \eta=\left(\begin{array}{cc}
    \eta_+& 0\\
    0 & \eta_- \end{array}\right),
    \label{t-e}\\
    \qq &=&\left(\begin{array}{cc}
    0 & 0 \\
    D & 0 \end{array}\right),~~~~
    H=\left(\begin{array}{cc}
    H_+ & 0\\
    0 & H_- \end{array}\right),
    \label{p-susy-2}
    \eea
where $\eta_\pm$ is a Hermitian linear automorphism of ${\cal H}_\pm$, $D:{\cal H}_+\to {\cal H}_-$ is a
linear operator, and
    \be
    H_+:=\frac{1}{2}\, D^\sharp D,~~~~
    H_-:= \frac{1}{2}\, D \,D^\sharp.
    \label{H-H}
    \end{equation}
Note that, by definition,  $\qq^\sharp=\eta^{-1}{\qq}^\dagger\eta$,
    \be
    D^\sharp=\eta_+^{-1}D^\dagger\eta_-,
    \label{D=}
    \end{equation}
and that $H_\pm:{\cal H}_\pm\to {\cal H}_\pm$ are $\eta_\pm$-pseudo-Hermitian Hamiltonians satisfying the
intertwining relations
    \be
    D\,H_+=H_-D,~~~~D^\sharp H_-=H_+ D^\sharp\;.
    \label{inter}
    \end{equation}
As a consequence, $H_+$ and $H_-$ are isospectral, $D$ maps the eigenvectors of $H_+$ to those of $H_-$
and $D^\sharp$ does the converse, except for those eigenvectors that are eliminated by these operators.
More specifically, suppose that $H_\pm$ has a complete biorthonormal eigenbasis
$\{|\psi_n^\pm,a\kt,|\phi_n^\pm,a\kt\}$ satisfying
    \[H_\pm|\psi_n^\pm,a\kt=E_n^\pm|\psi_n^\pm,a\kt,~~~~
    H^\dagger_\pm|\phi_n^\pm,a\kt=E_n^{\pm *} |\phi_n^\pm,a\kt.\]
Then, $D|\psi^+_n,a\kt$ is either zero in which case $E_n^+=0$, or it is an eigenvector of $H_-$ with
eigenvalue $E_n^+$; $D^\sharp|\psi^-_n,a\kt$ is either zero in which case $E_n^-=0$, or it is
an eigenvector of $H_+$ with eigenvalue $E_n^-$. Similarly $D^\dagger$ and $D^{\sharp\dagger}$ relate
the eigenvectors $|\phi_n^\pm,a\kt$ of $H_\pm^\dagger$.

An interesting situation arises when one of the automorphisms $\eta_\pm$ is trivial, e.g., $\eta_+=1$. In
this case, $H_+$ is a Hermitian Hamiltonian with a real spectrum, and pseudo-supersymmetry implies that
the pseudo-Hermitian Hamiltonian $H_-$ --- which is generally non-Hermitian --- must have a real spectrum
as well. This is not the only way to generate non-Hermitian Hamiltonians with a real spectrum. In the next
section we shall use pseudo-supersymmetry to construct a class of non-Hermitian Hamiltonians that have a
real spectrum.

\section{A Class of Non-Hermitian Hamiltonians with a Real Spectrum}

Consider the class of pseudo-supersymmetric systems corresponding to the choices:
    \bea
    &&{\cal H}_\pm={\cal H}=L^2(\R),~~~\eta_\pm=\pm P,
    \label{e10}\\
    &&D=p+f(x)+ig(x),
    \label{D}
    \eea
where $f$ and $g$ are real-valued functions. We can express these functions in the form
    \be
    f(x)=f_+(x)+f_-(x),~~~~g(x)=g_-(x)+g_+(x),
    \label{f-g}
    \end{equation}
where $f_+$ and $g_+$ are even functions of $x$, and $f_-$ and $g_-$ are odd functions. In view of
Eqs.~(\ref{e10}) -- (\ref{f-g}), (\ref{D=}), and (\ref{H-H}), we have
    \bea
    D^\sharp&=&p-f_+(x)+f_-(x)+i[g_+(x)-g_-(x)]\;,
    \label{D-s}\\
    H_\pm&=& \frac{1}{2}\left( [p+f_-(x)]^2+g'_-(x)\pm g_-^2-f_+^2-i[2g_-(x)f_+(x)\pm f'_+(x)]
    +K\right)\;,
    \label{H-pm-1=}\\
    K&:=&i\{g_+(x),p\}+g_+(x)[2if_-(x)-g_+(x)]\;,
    \label{K}
    \eea
where a prime means a derivative and $\{~,~\}$ stands for the anticommutator.

Next, we demand that $H_+$ is a Hermitian Hamiltonian. The
necessary and sufficient condition for the Hermiticity of $H_+$
and non-Hermiticity of $H_-$ is
    \be
    g_+(x)=0~~~~{\rm and}~~~~g_-(x) = -\frac{f_+'(x)}{2f_+(x)}.
    \label{gg}
    \end{equation}
Introducing the even function $\xi(x):=\ln |f_+(x)/\lambda|$ for
some $\lambda\in\R-\{0\}$, and using Eqs.~(\ref{H-pm-1=}) --
(\ref{gg}), we have
    \bea
    H_+&=& \frac{1}{2}\left( [p+f_-(x)]^2+\frac{1}{4}\,\xi'(x)^2-\frac{1}{2}\,\xi''(x)
    -\lambda^2\, e^{2\xi(x)}\right)\;,
    \label{H+}\\
    H_-&=& \frac{1}{2}\left( [p+f_-(x)]^2-\frac{1}{4}\,\xi'(x)^2-\frac{1}{2}\,\xi''(x)
    -\lambda^2\, e^{2\xi(x)}+2i\lambda\, e^{\xi(x)}\xi'(x) \right)\;.
    \label{H-}
    \eea
By construction, $H_\pm$ are pseudo-Hermitian pseudo-supersymmetric partners. In particular, they are
isospectral. $H_+$ happens to be a Hermitian operator. This implies that the eigenvalues of both
$H_+$ and $H_-$ are real. Furthermore, for $f_-(x)\ne 0$, $H_-$ is not $PT$-invariant. This is a
concrete example of a non-Hermitian Hamiltonian with a real spectrum that fails to be
$PT$-symmetric.

Eq.~(\ref{H-}) provides a large class of non-Hermitian Hamiltonians with a real spectrum whose members
are determined by the choice of functions $f_-$ and $\xi$. This class includes Hamiltonians with a discrete
spectrum. For example let $\xi(x)=-(x/\ell)^{2n}$, where $n$ is a
positive integer and $\ell$ is a positive real parameter with the dimension of length. Then
    \bea
    H_\pm&=&\frac{1}{2}\,[p+f_-(x)]^2+V_\pm(x)\;,\nn
    \\
    V_+&=&\frac{1}{2}\left( n^2\ell^{-4n}x^{4n-2}+n(2n-1)\ell^{-2n}x^{2n-2}
    -\lambda^2e^{-2\ell^{-2n}x^{2n}}\right)\;,\nn
    \\
    V_-&=&\frac{1}{2}\left(- n^2\ell^{-4n}x^{4n-2}+n(2n-1)\ell^{-2n}x^{2n-2}
    -\lambda^2e^{-2\ell^{-2n}x^{2n}}-4 i \lambda n\ell^{-2n}x^{2n-1}e^{-\ell^{-2n}x^{2n}}\right)
    \nn\;.
    \eea
It is not difficult to see that $H_+$ is a Hermitian Hamiltonian with a discrete spectrum. Therefore, $H_-$
has a real discrete spectrum as well.

\section{Conclusion}

In this article, we have introduced the concept of a
pseudo-Hermitian operator and showed that the desirable spectral
properties attributed to $PT$-symmetry are in fact consequences of
pseudo-Hermiticity of the corresponding Hamiltonians. We have
derived various properties of pseudo-Hermitian conjugation and
pseudo-Hermitian operators. In particular, we showed how the
defining automorphism $\eta$ is linked to the eigenvectors of an
$\eta$-pseudo Hermitian Hamiltonian $H$ with a complete
biorthonormal eigenbasis. As the corresponding eigenbasis is
subject to gauge transformations, the automorphism with respect to
which $H$ is pseudo-Hermitian is not unique. This raises the
question of the classification of the equivalence classes of
automorphisms that lead to the same notion of pseudo-Hermiticity
for a given Hamiltonian. We have given a brief discussion of this
problem and showed its connection with symmetries of the
Hamiltonian. We have also developed a generalization of
supersymmetry that would apply for general pseudo-Hermitian
Hamiltonians, and used it to construct a class of pseudo-Hermitian
Hamiltonians with a real spectrum.

A particularly interesting result of our investigations is that
all the $PT$-symmetric Hamiltonians that admit a complete
biorthonormal eigenbasis are pseudo-Hermitian. In this sense,
pseudo-Hermiticity is a generalization of $PT$-symmetry.

For a $PT$-symmetric Hamiltonian, the exactness of $PT$-symmetry
implies the reality of the spectrum. More specifically, if an
eigenvector $|E\kt$ is $PT$-invariant, $PT|E\kt=|E\kt$, then the
corresponding eigenvalue $E$ is real. A similar condition for a
general pseudo-Hermitian Hamiltonian is not known.
Pseudo-Hermiticity is only a necessary condition for the reality
of the spectrum, not a sufficient condition. In contrast,
$PT$-symmetry is neither necessary nor sufficient. The exact
$PT$-symmetry is a sufficient condition. But for a given
$PT$-symmetric Hamiltonian it is not easy to determine the
exactness of $PT$-symmetry without actually solving the
corresponding eigenvalue problem.

We hope that the concepts developed in this article provide the
material for a more rigorous study of the foundation of
pseudo-unitary quantum mechanics.

\section*{Acknowledgment}
I would like to thank M.~Znojil for bringing $PT$-symmetry to my
attention and the referee for his most invaluable comments. This
project was supported by the Young Researcher Award Program
(GEBIP) of the Turkish Academy of Sciences.

\ed